\documentclass[sigconf]{acmart} 
\usepackage[ruled,vlined,linesnumbered]{algorithm2e}
\usepackage{subcaption}
\usepackage{makecell}

\usepackage{multirow}
\usepackage{graphicx}
\AtBeginDocument{%
  }

\copyrightyear{2026}
\acmYear{2026}
\setcopyright{cc}
\setcctype{by}
\acmConference[KDD '26]{Proceedings of the 32nd ACM SIGKDD Conference on Knowledge Discovery and Data Mining V.2}{August 09--13, 2026}{Jeju Island, Republic of Korea}
\acmBooktitle{Proceedings of the 32nd ACM SIGKDD Conference on Knowledge Discovery and Data Mining V.2 (KDD '26), August 09--13, 2026, Jeju Island, Republic of Korea}
\acmDOI{10.1145/3770855.3818493}
\acmISBN{979-8-4007-2259-2/2026/08}

\begin{document}

\title{TMallGS: Scaling Unified Feature and Sequence Modeling for Generative E-commerce Search}

\author{Zhentao Song}
\email{220245077@seu.edu.cn}
\affiliation{%
  \institution{Southeast University}
  \city{Nanjing}
  \country{China}}

\author{Yufeng Gao}
\authornote{Yufeng Gao and Jing Wang are co-corresponding authors.}
\email{gyf265697@taobao.com}
\affiliation{%
  \institution{Taobao \& Tmall Group of Alibaba}
  \city{Hangzhou}
  \country{China}}

\author{Xing Fang}
\email{fangxing.fx@taobao.com}
\affiliation{%
  \institution{Taobao \& Tmall Group of Alibaba}
  \city{Hangzhou}
  \country{China}}

\author{Jing Wang}
\authornotemark[1]
\email{jing.wangj1@taobao.com}
\affiliation{%
  \institution{Taobao \& Tmall Group of Alibaba}
  \city{Hangzhou}
  \country{China}}

\author{Bokang Wang}
\email{wangbokang.wbk@alibaba-inc.com}
\affiliation{%
  \institution{Taobao \& Tmall Group of Alibaba}
  \city{Hangzhou}
  \country{China}}

\author{Guangxin Song}
\email{guangxin.sgx@alibaba-inc.com}
\affiliation{%
  \institution{Taobao \& Tmall Group of Alibaba}
  \city{Hangzhou}
  \country{China}}

\author{Yipin Dai}
\email{220245072@seu.edu.cn}
\affiliation{%
  \institution{Southeast University}
  \city{Nanjing}
  \country{China}}

\author{He Guo}
\email{2401210268@stu.pku.edu.cn}
\affiliation{%
  \institution{Peking University}
  \city{Beijing}
  \country{China}}

\renewcommand{\shortauthors}{Zhentao Song et al.}

\begin{abstract}
In large-scale industrial search and ranking systems, Click-Through Rate (CTR) prediction is undergoing a paradigm shift from traditional Deep Learning Recommendation Models (DLRM) toward unified, compute-intensive Transformer architectures. The primary motivation for this transition is to leverage Model FLOPs Utilization (MFU) to achieve predictable performance gains through Scaling Laws. However, existing scaling approaches like OneTrans and Climber, often adopt an all-in-tokenization strategy when directly migrating Large Language Model (LLM) architectures, which neglects the unique feature heterogeneity. We propose \textbf{TmallGS}, a high-performance, scalable universal ranking architecture tailored for the Tmall precision ranking domain. TmallGS introduces five core innovations:

(1) \textbf{Hierarchical Distribution-Calibrated Tokenization:} To bridge the heterogeneity gap, we propose a coarse-to-fine pipeline combining Field-wise Saliency Reweighting (FSR) and Distribution-Calibrated Projection (DCP) to project diverse features into optimized subspaces.
(2) \textbf{Field-Adaptive Gated Transformer Backbone:} We employ Per-Field QKV projections and a noise-adaptive gating mechanism to refine semantic interactions and suppress element-wise noise.
(3) \textbf{Decoupled FiLM Late Fusion:} To preserve high-frequency explicit signals, we utilize Feature-wise Linear Modulation (FiLM) to dynamically modulate backbone embeddings with explicit cross-features.
(4) \textbf{Context-Aware Bias Decoupling:} Addressing systemic biases beyond position, we incorporate a Context-Aware Bias Net that leverages deep global context to orthogonally decouple bias factors from genuine user intent.
(5) \textbf{Error-Aware Progressive Training:} We propose a dynamically weighted loss function based on hierarchical prediction errors, which enables adaptive hard-sample mining to improve model robustness.

Extensive offline experiments and online A/B tests conducted in the Tmall\footnote{Tmall is China's largest B2C e-commerce platform.} Search Ranking stage demonstrate that TmallGS significantly boosts training throughput while delivering substantial gains in both UCTCVR and GMV metrics.
\end{abstract}



\maketitle

\section{Introduction}

Ranking stands as the pivotal nexus connecting explicit user intent ({\itshape Query}) with massive item inventories. 
For the past decade, the industrial landscape has been dominated by the Deep Learning Recommendation Model (DLRM) paradigm, exemplified by DNN-based architectures like DeepFM\cite{guo2017deepfm} and DCNv2\cite{wang2021dcn}. These models rely on a dual-track design: massive embedding tables for memorization and shallow MLPs for feature interaction \cite{cheng2016wide}. While effective in the early deep learning era, DLRMs have recently collided with a formidable industrial compute wall. Due to their inherently memory-bound characteristics, DLRMs suffer from inefficient memory access patterns, leading to critically low Tensor Core utilization (Model FLOPs Utilization, MFU) on modern GPUs. Furthermore, they exhibit a saturation effect, where scaling parameters yields diminishing marginal returns in performance\cite{zhang2024wukong}.

Inspired by the transformative Scaling Laws observed in Large Language Models (LLMs) \cite{kaplan2020scaling}, the industry is undergoing a fundamental paradigm shift. Recent pioneers have begun exploring compute-intensive architectures anchored by Transformer backbones \cite{vaswani2017attention, zhai2024actions}. By stacking deep Transformer layers, these models not only maximize hardware efficiency\cite{dao2022flashattention}  but also unlock superior capabilities for modeling ultra-long user behavior sequences\cite{pi2019practice, song2025lrea}. Despite the Transformer's dominance in general sequence modeling, adapting the LLM-style "All-in-One" architecture to high-precision search ranking introduces severe theoretical and practical defects\cite{zhang2026onetrans, huang2026hyformer,zhang2022revisiting}.

Unlike Recommendation Systems which tolerate fuzzy discovery, Search is strictly governed by Query Constraints. We identify three critical challenges that impede current scaling efforts: 
(1) \textbf{The Heterogeneity Gap:} Search inputs vary significantly, from short queries to long-tail user behaviors\cite{yu2025hhft, gui2023hiformer}. Forcing a unified attention mechanism across these diverse modalities triggers Gradient Conflicts. 
(2) \textbf{Signal Dilution:} Explicit cross-features (e.g., query-item matches) are decisive in search\cite{bian2022can}. However, deep Self-Attention inherently acts as a low-pass filter, over-smoothing these high-frequency hard signals. 
(3) \textbf{Optimization Instability:} Unlike the dense supervision in LLMs, CTR prediction relies on sparse binary feedback, leading to severe gradient vanishing in deep networks\cite{zong2025recis}.

To bridge these gaps, we propose TmallGS, a highly efficient architecture tailored for search scaling. Adhering to the philosophy of "Semantic Divide-and-Conquer, Interaction Decoupling," our contributions are summarized as follows:

(1) \textbf{Hierarchical Distribution-Calibrated Tokenization:} We propose a coarse-to-fine tokenization pipeline comprising Field-wise Saliency Reweighting (FSR) and Distribution-Calibrated Projection (DCP). This design projects heterogeneous features into calibrated subspaces, effectively solving the heterogeneity gap at the input level\cite{yu2025hhft}.

(2) \textbf{Field-Adaptive Gated Backbone:} We design a specialized Transformer backbone equipped with Per-Field QKV projections and Noise-Adaptive Gating\cite{qiu2026gated}. This mechanism refines semantic interactions by dynamically filtering noise from long behavior sequences\cite{zhou2019deep}.

(3) \textbf{Signal-Preserving Decoupling:} We introduce Decoupled FiLM Late Fusion\cite{perez2018film} to resolve the signal dilution problem. By explicitly modeling heavy cross-features via a modulation pathway, we ensure that high-frequency matching constraints are preserved alongside deep semantic abstractions.

(4) \textbf{Context-Aware Bias Net:} We design a Context-Aware Bias Net to address systemic biases. Unlike shallow position debiasing, this module leverages the deep global context anchor\cite{dai2025onepiece} to orthogonally decouple complex bias factors from semantic relevance\cite{yu2026transun}, optimizing GAUC directly.

(5) \textbf{Robust Progressive Optimization:} We introduce an Error-Aware Progressive Training strategy\cite{zong2025recis} to stabilize deep training via adaptive curriculum learning. Coupled with a Two-Stage Warm-up protocol, TmallGS achieves state-of-the-art performance and stability in both offline benchmarks and online A/B testing at Tmall APP Search.

\section{Related Work}
\textbf{Traditional Recommendation Paradigms.}
For the past decade, DLRMs paradigm has dominated industrial ranking. Pioneering works like Wide\&Deep\cite{cheng2016wide} and DeepFM\cite{guo2017deepfm} utilized shallow MLPs to capture feature interactions, while DCN-V2\cite{wang2021dcn} introduced explicit vector-wise cross networks. 
Other feature-interaction models, such as xDeepFM, AutoInt, and FiBiNET, further explore explicit high-order interactions, self-attentive feature interaction, and feature-importance weighting for CTR prediction~\cite{lian2018xdeepfm,song2019autoint,huang2019fibinet}. To incorporate user history, DIN\cite{zhou2018deep} and DIEN\cite{zhou2019deep} proposed Target Attention mechanisms to weigh historical behaviors against candidate items, while BST \cite{chen2019behavior} and DSIN \cite{feng2019deep} explored Transformer and session-based modeling. However, the fundamental bottleneck of DLRM lies in its Memory-Bound nature. These models rely heavily on massive embedding tables for memorization\cite{ma2018entire}, resulting in sparse, irregular memory access patterns that critically underutilize modern GPU Tensor Cores, which lead to low MFU. Furthermore, DLRMs face a Scaling Ceiling\cite{zhang2024wukong}: simply widening MLPs or deepening attention layers yields diminishing marginal returns and fails to efficiently model ultra-long sequences (e.g., lifetime history) due to memory bandwidth constraints\cite{pi2019practice}. This "Industrial Compute Wall" necessitates a paradigm shift toward compute-intensive architectures that can leverage Scaling Laws.
\begin{figure*}[t]
    \centering

    \includegraphics[width=1\textwidth]{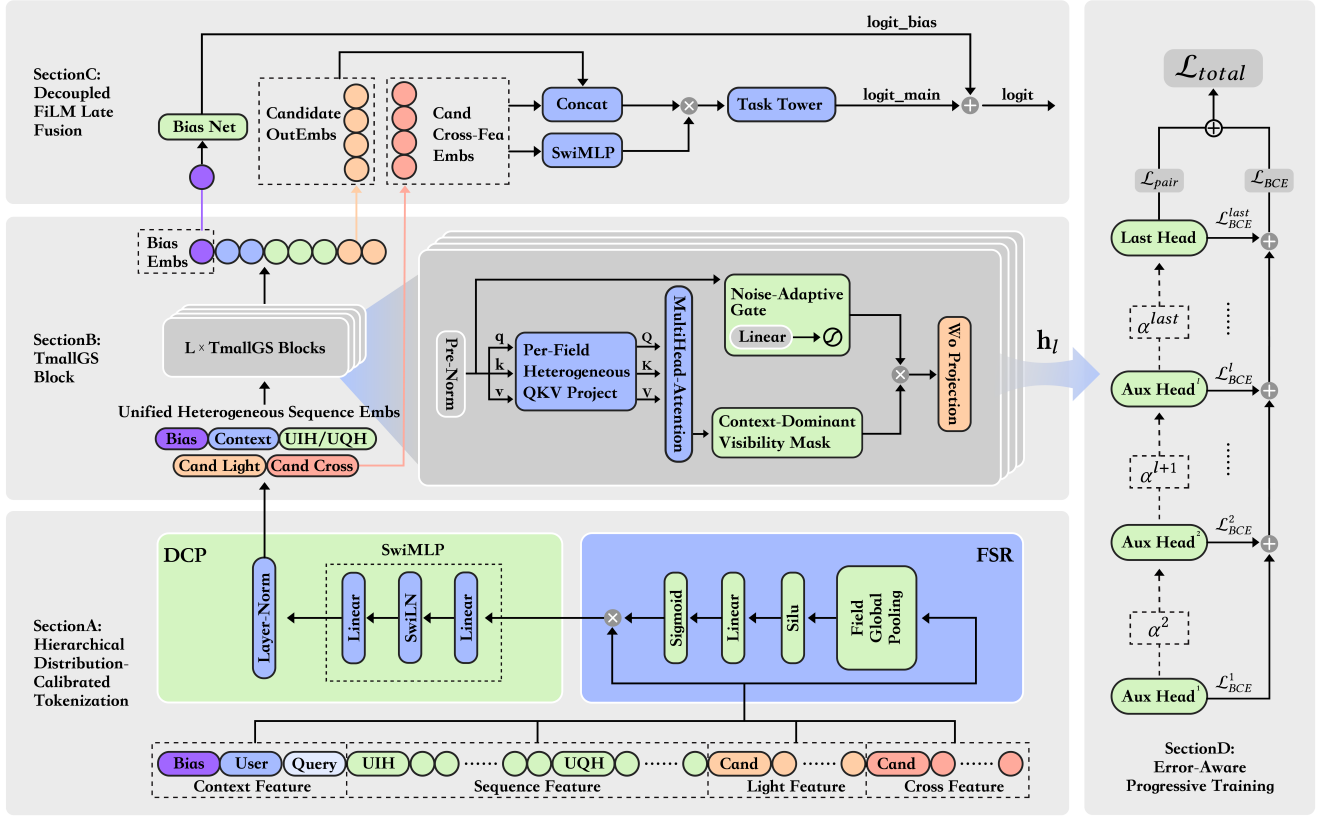} 
    \caption{The overall architecture of TmallGS. (Left) The distinct processing flow comprising three stages: \textbf{Section A} uses FSR and DCP for hierarchical tokenization; \textbf{Section B} employs the TmallGS Block with Per-Field QKV and Noise-Adaptive Gating; \textbf{Section C} performs Decoupled FiLM Late Fusion to recover explicit signals. (Right) The Error-Aware Progressive Training strategy with auxiliary supervision.}
    \label{fig:architecture}
\end{figure*}

\textbf{Scaling Transformers.}
Inspired by LLM Scaling Laws, the industry has pivoted towards Unified Transformer Backbones to maximize compute efficiency\cite{zhai2024actions, zhu2025rankmixer}. OneTrans\cite{zhang2026onetrans} utilizes a unified tokenizer for all-layer interaction via a shared Transformer. While theoretically capturing global correlations, tokenizing strong explicit signals (e.g., matching scores) risks Signal Dilution due to the low-pass filtering nature of deep Self-Attention.  HyFormer\cite{huang2026hyformer} attempted to mitigate this via asymmetric interaction. OnePiece\cite{dai2025onepiece} introduced Reasoning Blocks to unify search contexts, but its high computational complexity poses challenges for low-latency industrial deployment. Nevertheless, most existing unified architectures lack Architectural Decoupling. By forcing statistical features and semantic tokens into a homogeneous attention stream, these models often suffer from a lack of Inductive Bias required for precise relevance matching\cite{bian2022can} and struggle to balance the gradient conflicts between sparse features and dense semantic representations\cite{yan2022apg}.

\textbf{SID Paradigm GRs.}
The Generative Retrieval (or SID) paradigm, represented by TIGER\cite{rajput2023recommender}, and recently OneRec\cite{deng2025onerec}, OneSearch\cite{chen2025onesearch}\cite{chen2026onesearch}, reframes recommendation as end-to-end ID generation. While promising for recall\cite{liang2025tbgrecall, han2025mtgr}, this black-box approach faces fundamental Controllability Challenges in industrial ranking. Specifically, end-to-end generation bypasses critical Intermediate Strategic Interventions (e.g., inventory filtering, ad insertion) essential for commercial systems. Furthermore, generative models inherently rely on probabilistic sampling, making it difficult to strictly adhere to Hard Query Constraints (e.g., exact attribute matching), leading to potential hallucinations that are unacceptable in high-precision search scenarios.

\section{Preliminaries}
\noindent \textbf{Problem Formulation.} We formulate the CTR prediction task in industrial search ranking as learning a probability estimator $f_{\theta}: \mathcal{X} \to [0, 1]$. Let the training corpus be denoted as $\mathcal{S} = \{(u_i, q_i, \mathcal{C}_i, \mathbf{y}_i)\}_{i=1}^N$, where $u_i$ represents the user context, $q_i$ is the search query, and $\mathcal{C}_i = \{c_1, \dots, c_M\}$ is the set of candidate items in the session. The label vector $\mathbf{y}_i \in \{0, 1\}^M$ indicates the click status for each candidate. 

The heterogeneous feature space $\mathcal{X}$ is partitioned into three logical subspaces to support our decoupled architecture: (1) \textbf{Context Features ($\mathbf{x}_{ctx}$)} encompass request-level global priors shared across all candidates, such as query text, user profile, and pagination biases; (2) \textbf{Sequence Features ($\mathbf{x}_{seq}$)} consist of variable-length historical behaviors, including User Interaction History ($\mathcal{H}_{u}$) and Query History ($\mathcal{H}_{q}$), which capture long-term intent shifts; and (3) \textbf{Candidate Features ($\mathbf{x}_{cand}$)} involve item-specific attributes. Crucially, we bifurcate these into \textit{Light Attributes} $\mathbf{x}_{c}^{light}$ (e.g., ItemID) for semantic representation learning, and \textit{Heavy Cross-Attributes} $\mathbf{x}_{c}^{heavy}$ (e.g., Q2I matching scores) for precise calibration.

\section{Methodology}
\subsection{Architecture Overview}
We propose \textbf{TmallGS}, a scalable generative search ranking architecture designed to bridge the \textit{Heterogeneity Gap} and prevent \textit{Signal Dilution}. As illustrated in Figure \ref{fig:architecture}, TmallGS departs from the monolithic all-in-one tokenization used in LLMs. Instead, it adopts a semantic-interaction decoupling paradigm comprising five synergistic modules:
(1) \textbf{Hierarchical Distribution-Calibrated Tokenization} that projects heterogeneous features into optimized subspaces;
(2) \textbf{Field-Adaptive Gated Transformer Backbone} which employs \textit{Per-Field QKV Projections} and \textit{Noise-Adaptive Gating} to refine semantic interactions;
(3) \textbf{Decoupled FiLM Late Fusion} to recover high-frequency explicit signals;
(4) \textbf{Context-Aware Bias Net} for explicitly modeling and monitoring context biases;
(5) \textbf{Error-Aware Progressive Training} to enforce a robust curriculum learning capability.

\subsection{Hierarchical Distribution-Calibrated Tokenization}
Tokenization serves as the foundational bridge between sparse raw features and dense semantic interactions. Given the massive scale and noise inherent in industrial feature vocabularies, a simple linear projection is insufficient\cite{yan2022apg}. Conversely, standard heavy gating (e.g., SwiGLU-FFN) incurs prohibitive computational costs\cite{dao2022flashattention}.
To resolve this, we propose a \textbf{Coarse-to-Fine Tokenization Pipeline} that acts as a progressive denoising filter. This pipeline consists of two coupled stages: \textit{Field-wise Saliency Reweighting} for macroscopic noise suppression, followed by \textit{Distribution-Calibrated Projection} for microscopic feature adaptation.

\subsubsection{\textbf{Stage 1: Field-wise Saliency Reweighting (FSR)}}
Before mapping raw features into the latent space, it is critical to calibrate the importance of different feature fields based on the global context\cite{dai2025onepiece}. We introduce a lightweight \textbf{Field-wise Saliency Reweighting} module to adaptively re-weight input fields.

Let $\mathcal{X} = \{\mathbf{X}_1, \dots, \mathbf{X}_F\}$ denote the input feature fields, where $\mathbf{X}_i \in \mathbb{R}^{L \times d_i}$ represents the $i$-th field with sequence length $L$ and raw dimension $d_i$.
First, we perform \textbf{Global Context Aggregation} to capture the macroscopic intensity of each field. Unlike standard SE-blocks that only pool the channel dimension, we aggregate over both the temporal ($L$) and feature ($d_i$) dimensions to obtain a robust global descriptor $z_i$:
\begin{equation}
    z_i = \frac{1}{L \cdot d_i} \sum_{t=1}^{L} \sum_{k=1}^{d_i} \mathbf{X}_{i,t,k}, \quad \mathbf{z} = [z_1, \dots, z_F] \in \mathbb{R}^F.
\end{equation}
Next, a gating network models the inter-field dependencies to generate saliency weights $\mathbf{w} \in \mathbb{R}^F$:
\begin{equation}
    \mathbf{w} = \sigma(\mathbf{W}_{ex} \cdot \delta(\mathbf{W}_{sq} \mathbf{z})),
\end{equation}
where $\delta$ is the SiLU activation, and $\mathbf{W}_{sq}, \mathbf{W}_{ex}$ form a bottleneck structure to capture field correlations efficiently. The raw features are then dynamically modulated: $\tilde{\mathbf{X}}_i = w_i \cdot \mathbf{X}_i$. This step acts as a soft field selector, amplifying globally informative fields while dampening irrelevant ones prior to detailed projection.

\subsubsection{\textbf{Stage 2: Distribution-Calibrated Projection (DCP)}}
Following the coarse-grained reweighting, we employ the \textbf{Distribution-Calibrated Projection} to project the modulated features into the semantic space.
We introduce the \textit{Calibrated-Swish} activation ($\phi_{cal}$), which implements a self-gating mechanism driven by layer statistics\cite{qiu2026gated}. For the reweighted field $\tilde{\mathbf{X}}_i$, the token embedding $\mathbf{E}_i$ is computed as:
\begin{equation}
    \phi_{cal}(\mathbf{H}) = \mathbf{H} \odot \sigma(\text{LayerNorm}(\mathbf{H})),
\end{equation}
\begin{equation}
    \mathbf{E}_i = \text{MLP}_{cal}(\tilde{\mathbf{X}}_i) = \left( \phi_{cal}(\tilde{\mathbf{X}}_i \mathbf{W}_1 + \mathbf{b}_1) \right) \mathbf{W}_2.
\end{equation}
Here, Layer Normalization standardizes the latent distribution, ensuring the self-gating signal is strictly centered within the gradient-sensitive region.
\textbf{Efficiency Analysis:} This two-stage design provides a favorable trade-off: FSR introduces negligible overhead (operating on pooled scalars), while DCP reduces projection FLOPs by about 33\% compared to SwiGLU-FFN\cite{dao2022flashattention}. Together, they achieve robust non-linear modeling with high inference throughput.

\subsection{Unified Heterogeneous Sequence Construction}
To fully leverage the global receptive field of the Transformer\cite{vaswani2017attention, zhai2024actions}, we propose a \textbf{Request-Level Sequence Construction} strategy. Unlike item-centric paradigms that redundantly encode context for each candidate, we align the sequence strictly with the User-Query Request, constructing a unified input $\mathbf{E}_{in}$ shared across all candidates\cite{zhang2026onetrans}.

\paragraph{\textbf{Tokenization \& Assembly}}
We define five core semantic components: (1) \textbf{Context Bias Anchor ($\mathbf{e}_{bias}$)} tokenizes the pagination index to aggregate global context\cite{yu2026transun}; (2) \textbf{Context \& Query ($\mathbf{e}_{ctx}, \mathbf{e}_{qry}$)} represent user profile and current intent; (3) \textbf{User History ($\mathbf{H}_{uih}, \mathbf{H}_{uqh}$)} captures interactions and explicit intent shifts\cite{zhou2019deep}, where query tokens aggregate linguistic features; (4) \textbf{Candidate ($\mathbf{e}_{cand}$)} is restricted to \textit{Light Attributes} (e.g., ItemID) to align with history, bypassing heavy cross-attributes. 
The final sequence is positioned with the Anchor at the start to ensure global attention coverage:
\begin{equation}
    \mathbf{E}_{in} = [\mathbf{e}_{bias}; \mathbf{e}_{ctx}; \mathbf{e}_{qry}; \mathbf{H}_{uih}; \mathbf{H}_{uqh}; \mathbf{e}_{cand}^{light}].
\end{equation}
This unified sequence is fed into the backbone, where field-specific QKV projections learn distinct interaction patterns\cite{gui2023hiformer}.

\subsection{TmallGS Block}
The core reasoning engine consists of $L$ stacked TmallGS Blocks. To handle the unique intra-session ranking structure and the noise inherent in behavioral data, we redesign the standard Transformer block\cite{vaswani2017attention} with field-specific projections, a permutation-invariant visibility mask, and a noise-adaptive gating mechanism\cite{qiu2026gated}.

\subsubsection{\textbf{Per-Field Heterogeneous QKV Projection}}
Standard Transformers project all tokens into a shared latent space, which neglects the manifold discrepancies among search fields\cite{yu2025hhft}. To resolve this, we partition the unified input sequence $\mathbf{E}_{in}$ into field-specific segments. We employ \textbf{Per-Field Projections}\cite{gui2023hiformer}, where each field $\phi$ (e.g., Query, UIH, Candidate) utilizes a unique set of projection matrices. For the $i$-th token $\mathbf{h}_i$ belonging to field $\phi(i)$, the query, key, and value vectors are computed as:
\begin{equation}
    \mathbf{q}_i = \mathbf{h}_i \mathbf{W}_Q^{\phi(i)}, \quad \mathbf{k}_i = \mathbf{h}_i \mathbf{W}_K^{\phi(i)}, \quad \mathbf{v}_i = \mathbf{h}_i \mathbf{W}_V^{\phi(i)}.
\end{equation}
This design acts as a manifold alignment mechanism, mitigating gradient conflicts by allowing distinct semantic rotations for heterogeneous modalities\cite{yan2022apg}.

\subsubsection{\textbf{Context-Dominant Visibility Mask}}
In search ranking, multiple candidate items are scored simultaneously within a single batch. To ensure efficiency and theoretical correctness, we impose a strict Attention Mask $\mathbf{M} \in \mathbb{R}^{T \times T}$.
Let the input sequence be divided into a \textit{Context Prefix} set $\Omega_{ctx} = \{Anchor, User, Query, UQH, UIH\}$ and a \textit{Candidate} set $\Omega_{cand}$. The visibility rules are defined as:
\begin{itemize}
    \item \textbf{Context Full-Attention:} Tokens within $\Omega_{ctx}$ are fully visible to each other to model deep global context interactions\cite{dai2025onepiece}.
    \item \textbf{Candidate Independence:} Each candidate token $c_i \in \Omega_{cand}$ can attend to all tokens in $\Omega_{ctx}$ and itself, but is strictly invisible to other candidates $c_j (j \neq i)$.
\end{itemize}
Formally, the mask is defined as:
\begin{equation}
    \mathbf{M}_{ij} = 
    \begin{cases} 
    0 & \text{if } j \in \Omega_{ctx} \text{ or } i = j \\
    -\infty & \text{if } i, j \in \Omega_{cand} \text{ and } i \neq j 
    \end{cases}
\end{equation}
This masking strategy prevents information leakage between candidates (shortcut learning) and ensures that the ranking score of a candidate depends solely on its interaction with the context, not on the other items in the same page.

\subsubsection{\textbf{Noise-Adaptive Gated Attention}}
User behavior sequences often contain noise (e.g., accidental clicks) irrelevant to the current query\cite{zhou2018deep}. Standard Softmax normalization forces the model to allocate attention weights summing to 1, causing the noise propagation problem where irrelevant history tokens are assigned non-trivial weights.
To address this, we introduce a gating mechanism inserted \textit{after} the attention operation but \textit{before} the output projection $\mathbf{W}_O$\cite{qiu2026gated}.

Crucially, the gate is generated from the \textbf{Pre-Norm} output $\mathbf{X}_{norm}$ rather than the unstable attention output. Since $\mathbf{X}_{norm}$ retains the stable residual identity, it provides a robust reference for determining feature saliency. The computation flow is:
\begin{align}
\mathbf{A} 
&= \operatorname{Attention}(\mathbf{Q}, \mathbf{K}, \mathbf{V}), \\
\mathbf{G} 
&= \sigma(\mathbf{X}_{\mathrm{norm}}\mathbf{W}_{\mathrm{gate}} + \mathbf{b}_{\mathrm{gate}}), \\
\mathbf{H}
&= (\mathbf{A} \odot \mathbf{G})\mathbf{W}_{O}
\end{align}
where $\odot$ denotes element-wise multiplication and $\mathbf{A}$ is the standard attention output. By multiplying with $\mathbf{G} \in (0, 1)$, we empower the attention mechanism with an un-normalized capability. When the context is irrelevant, the gate $\mathbf{G}$ can saturate to 0, dynamically suppressing the noise propagation. This effectively breaks the linear bottleneck of the standard Value-Projection path ($W_V \cdot W_O$), enhancing the model's non-linear capacity to capture complex matching patterns\cite{bian2022can, yan2022apg}.

\subsection{Decoupled FiLM Late Fusion}
Deep Transformer layers act as low-pass filters, which can over-smooth high-frequency signals such as exact string matches or statistical interaction rates found in $\mathbf{x}_{c}^{heavy}$\cite{zhang2026onetrans}. To preserve these signals, we employ a Feature-wise Linear Modulation (FiLM)\cite{perez2018film} mechanism at the final output stage.

Let $\mathbf{h}_{c}^{L}$ denote the semantic output of the final Transformer block for candidate $c$, and $\mathbf{e}_{c}^{all}$ denote the comprehensive embedding of the candidate (combining both light and heavy features). We first concatenate them to form a unified representation basis:
\begin{equation}
    \mathbf{u}_{c} = [\mathbf{h}_{c}^{L} \oplus \mathbf{e}_{c}^{all}].
\end{equation}
We then employ a specific Task Tower (MLP) that takes the candidate's comprehensive features $\mathbf{e}_{c}^{all}$ as input to generate affine transformation coefficients:
\begin{equation}
    \boldsymbol{\gamma}_c, \boldsymbol{\beta}_c = \text{MLP}_{FiLM}(\mathbf{e}_{c}^{all}), \quad \text{where } \boldsymbol{\gamma}_c, \boldsymbol{\beta}_c \in \mathbb{R}^{d_{u}}.
\end{equation}
The final representation is obtained by dynamically scaling and shifting the unified vector:
\begin{equation}
    \mathbf{v}_{final} = (1 + \boldsymbol{\gamma}_c) \odot \mathbf{u}_{c} + \boldsymbol{\beta}_c.
\end{equation}
This mechanism allows the explicit cross-features to sharply modulate the semantic representations, recovering signal resolution lost during deep propagation. The main semantic logit is then computed as\cite{guo2017deepfm}:
\begin{equation}
    \text{logit}_{main} = \text{MLP}_{head}(\mathbf{v}_{final}).
\end{equation}

\subsection{Context-Aware Bias Net}
In Tmall search, one request corresponds to one displayed page of candidate items. 
When the user scrolls to the next page, the system creates a new request and a new training instance. 
Therefore, all candidates within the same request share request-level factors such as pagination, query context, user state, and traffic conditions. 
We introduce a Context Anchor token to summarize these shared factors.
To decouple systemic biases from semantic relevance, we propose a \textbf{Context-Aware Bias Net} that leverages the deep contextual encoding of the Context Anchor token\cite{yu2026transun}.
Unlike shallow bias towers, we extract the final hidden state of the Context Anchor, $\mathbf{h}_{bias}^{L}$, which has attended to the entire request context\cite{dai2025onepiece}. This vector is projected into a scalar bias term and additively fused with the main semantic logit $\text{logit}_{main}$ derived in Section 4.5:
\begin{equation}
    \text{logit}_{bias} = \text{SwiMLP}_{bias}(\mathbf{h}_{bias}^{L}), \quad \hat{y} = \sigma(\text{logit}_{main} + \text{logit}_{bias}).
\end{equation}

Crucially, this additive formulation enables an orthogonal optimization strategy. When computing the Pairwise Loss between a clicked item $c^+$ and a non-clicked item $c^-$ within the same request, the context-level bias term naturally cancels out:
\begin{equation}
\begin{split}
    \Delta \text{logit} &= (\text{logit}_{main}^+ + \text{logit}_{bias}) - (\text{logit}_{main}^- + \text{logit}_{bias}) \\
    &\equiv \text{logit}_{main}^+ - \text{logit}_{main}^-.
\end{split}
\end{equation}
This cancellation creates a clean separation of concerns: the Bias Net absorbs global click priors to stabilize AUC, while the Pairwise Loss is relieved from modeling systemic bias and focuses purely on relative relevance ($\text{logit}_{main}$), directly boosting GAUC. This prevents the calibration degradation typically associated with aggressive pairwise optimization\cite{zong2025recis}.

\subsection{Error-Aware Progressive Training}
To mitigate the vanishing gradient problem in deep networks and facilitate \textit{Curriculum Learning}, we propose a dynamic Error-Aware Progressive Training strategy.
Instead of treating all samples equally, we utilize the prediction error of the $l$-th layer to supervise the $(l+1)$-th layer, forcing deeper layers to focus on hard samples misclassified by shallower layers.

Specifically, at each layer $l$, we attach a lightweight auxiliary head to the candidate token output $\mathbf{h}_{c}^{l}$ to generate an intermediate prediction $\hat{y}^l$. We calculate the absolute prediction error (without clipping) as the hardness metric:
\begin{equation}
    \delta^l = |y - \hat{y}^l|.
\end{equation}
This error determines the loss weight $\alpha^{l+1}$ for the \textit{next} layer:
\begin{equation}
    \alpha^{l+1} = 1 + \lambda \cdot \text{StopGradient}(\delta^l).
\end{equation}
The \texttt{StopGradient} operator ensures that the weight $\alpha$ acts purely as a scalar modulator, preventing unstable gradients from flowing back through the weighting mechanism itself.

\textbf{Total Objective.} Finally, to explicitly optimize the intra-session ranking order (GAUC) alongside the progressive calibration, we incorporate a pairwise constraint on the final output:
\begin{equation}
    \mathcal{L}_{pair} = - \frac{1}{|\mathcal{P}|} \sum_{(c^+, c^-) \in \mathcal{P}} \log \sigma \left( \hat{y}(c^+) - \hat{y}(c^-) \right).
\end{equation}
The total objective is a weighted summation:
\begin{equation}
    \mathcal{L}_{total} = \underbrace{\sum_{l=1}^{L} \alpha^{l} \mathcal{L}_{BCE}(\hat{y}^l, y)}_{\text{Progressive Pointwise}} + \underbrace{\gamma \cdot \mathcal{L}_{pair}(\hat{y}^L)}_{\text{Pairwise Ranking}}
\end{equation}
where $\mathcal{P}$ denotes the set of valid training pairs constructed within each query session\cite{ma2018entire}, with $c^+$ and $c^-$ representing the clicked and non-clicked candidates respectively, and $|\mathcal{P}|$ is the total pair count. $\gamma$ is a hyperparameter that balances the BCE and Pairwise objectives. This hybrid design ensures robust convergence via deep supervision\cite{zong2025recis} while directly boosting the ranking capability tailored for search.

\section{Experiments}
In this section, we conduct extensive experiments to answer the following research questions:
\begin{itemize}
    \item \textbf{RQ1 (Effectiveness):} Does TmallGS outperform state-of-the-art industrial ranking models?
    \item \textbf{RQ2 (Ablation):} What is the contribution of each proposed component to the final performance?
    \item \textbf{RQ3 (Scalability):} Can TmallGS break the diminishing returns bottleneck of DLRMs and follow the scaling laws consistent with LLMs\cite{kaplan2020scaling}?
    \item \textbf{RQ4 (Efficiency):} Does the proposed architecture maintain a favorable trade-off between inference latency and model capacity for online deployment?
\end{itemize}

\begin{table}[h!]
    \centering
    \caption{Statistics of the Tmall Search Dataset.}
    \label{tab:dataset_stat}
    \vspace{-0.2cm} 
    \begin{tabular}{l|c}
        \toprule
        \textbf{Item} & \textbf{Statistics} \\
        \midrule
        Time Span & 31 Days \\
        Total Samples & $\sim$ 500 Million \\
        \midrule
        Avg. Sequence Length & 1,500 \\ 
        \midrule
        \# Users & \textit{13 Million} \\
        \# Items & \textit{74 Million} \\
        \midrule
        Training Set (Days 1-30) & \textit{484 Million} \\
        Test Set (Day 31) & \textit{16 Million} \\
        \bottomrule
    \end{tabular}
    \vspace{-0.5cm} 
\end{table}
\subsection{Experimental Setup}

\subsubsection{\textbf{Dataset.}}
Due to the absence of public large-scale search datasets that contain both ultra-long user behavior sequences and explicit query-item matching signals, standard benchmarks are insufficient for evaluating the long-context modeling and signal preservation capabilities of TmallGS. Consequently, we conduct experiments exclusively on a comprehensive industrial dataset collected from the Tmall App Search logs.

As summarized in Table \ref{tab:dataset_stat}, the dataset spans a consecutive period of 31 days, comprising over 500 million samples. To fully leverage the Transformer's\cite{vaswani2017attention} receptive field, we construct user behavior sequences with an average length of 1500 by truncating extended lifetime histories to the most recent interactions. This consistently long context setting covers a significantly broader history than typical industrial settings (usually under 200), posing a substantial challenge for efficient modeling\cite{zhai2024actions}. Strictly following the temporal order, we utilize the data from the first 30 days for training and the data from the last day for testing.

\subsubsection{\textbf{Baselines}}
To rigorously evaluate the performance of TmallGS, we compare it against a comprehensive set of competitive methods, categorized into traditional industrial paradigms and recent scalable architectures.

\textbf{Group 1: Traditional Models.}
We compare against standard DLRM baselines:
\begin{itemize}
    \item \textbf{DNN\cite{cheng2016wide}:} Uses a standard Embedding-MLP architecture to capture implicit feature interactions.
    \item \textbf{DIN\cite{zhou2018deep}:} Introduces local activation units to capture user interests; combined with RankMixer as our production baseline.
    \item \textbf{DCNv2\cite{wang2021dcn}:} Utilizes low-rank cross layers to efficiently learn bounded-degree explicit feature interactions.
    \item \textbf{APG\cite{yan2022apg}:} Dynamically generates instance-specific parameters via a hyper-network to handle diverse data distributions.
\end{itemize}

\textbf{Group 2: Scalable Architectures.}
These methods leverage Transformer backbones for scaling:
\begin{itemize}
    \item \textbf{HiFormer\cite{gui2023hiformer}:} Efficiently models long user behaviors via a two-stage hierarchical aggregation mechanism.
    \item \textbf{RankMixer\cite{zhu2025rankmixer}:} Replaces self-attention with multi-head token mixing to maximize GPU Model FLOPs Utilization (MFU).
    \item \textbf{HHFT\cite{yu2025hhft}:} Partitions heterogeneous features into semantic blocks with independent projections to prevent interference.
    \item \textbf{HSTU\cite{zhai2024actions}:} Replaces softmax-attention with point-wise aggregations to enable high-efficiency large-scale scaling.
    \item \textbf{MTGR\cite{han2025mtgr}:} Reframes CTR prediction as a sequence generation task, though often struggling with ranking precision.
    \item \textbf{OneTrans\cite{zhang2026onetrans}:} Serializes all context and sequence features into a single token stream for unified causal modeling.
\end{itemize}

\subsubsection{\textbf{Evaluation Metrics}}
We employ widely accepted metrics for industrial ranking evaluation to assess the performance of our proposed method: AUC and GAUC\cite{zhou2018deep, ma2018entire}.

The AUC metric evaluates the global ranking capability by measuring the probability that a randomly selected positive sample ranks higher than a randomly selected negative one. Let $\mathcal{D}^+$ and $\mathcal{D}^-$ denote the sets of positive and negative samples respectively. It is formally defined as:
\begin{equation}
    \text{AUC} = \frac{1}{|\mathcal{D}^+| |\mathcal{D}^-|} \sum_{i \in \mathcal{D}^+} \sum_{j \in \mathcal{D}^-} \mathbb{I}(\hat{y}_i > \hat{y}_j),
\end{equation}
where $\mathbb{I}(\cdot)$ is the indicator function which equals 1 if $\hat{y}_i > \hat{y}_j$ and 0 otherwise.

However, in search scenarios, the relative order of items within the same query session is more critical than global absolute scores. Therefore, we primarily rely on GAUC, which calculates the weighted average of AUC within each user-query group:
\begin{equation}
    \text{GAUC} = \frac{\sum_{(u,q)} w_{(u,q)} \times \text{AUC}_{(u,q)}}{\sum_{(u,q)} w_{(u,q)}},
\end{equation}
where $\text{AUC}_{(u,q)}$ is the AUC calculated strictly within the candidate set of session $(u,q)$, and $w_{(u,q)}$ denotes the number of impressions in that session. Note that in large-scale industrial recommendation, an absolute increase of 0.001 in GAUC is considered statistically significant.

\subsubsection{\textbf{Implementation Details}}
All models are implemented using a distributed training framework\cite{zong2025recis} and trained on a cluster of 16 high-performance Nvidia GPUs. The global batch size is set to 4096. learning rate is set to $2 \times 10^{-4}$.
Given the distinct convergence properties of sparse embeddings and dense Transformer parameters, we employ a Split-Optimizer strategy: For sparse model, we use the Adam optimizer to handle the non-stationary distribution of sparse features, ensuring adaptive learning rates for long-tail IDs;
For dense model, we use the AdamW optimizer for the TmallGS backbone and MLP towers. The decoupled weight decay\cite{hoffmann2022training} in AdamW effectively prevents overfitting in deep compute-intensive layers.
To ensure a fair comparison in our experiments, particularly in ablation studies, we standardize the model capacity: all DNN-based models utilize identical MLP hidden dimensions, and all Transformer-based architectures are configured with the same hidden dimension size.

\subsection{RQ1: Overall Performance}
We present the comparative results of the State-of-the-Art models on the Tmall Search Dataset in Table \ref{tab:main_results}. The models are evaluated based on AUC and GAUC. The relative improvement is calculated based on AUC and GAUC against baseline DIN. And our production baseline is DNN with Rankmixer.
\begin{table}[t]
    \centering
    \caption{Performance comparison on the Tmall Search Dataset. We report the relative improvement ($\Delta$) in percentage over the production baseline (DIN). \textbf{Bold} indicates the best performance, and \underline{underline} indicates the second best.}
    \label{tab:main_results}
    \renewcommand{\arraystretch}{1} 
    \setlength{\tabcolsep}{10pt} 
    \begin{tabular}{l|cc}
        \toprule
        \textbf{Model} & \textbf{$\Delta$ AUC (\%)} & \textbf{$\Delta$ GAUC (\%)} \\
        \midrule
        \multicolumn{3}{l}{\textit{Group 1: Traditional Models}} \\
        \midrule
        DNN-based                    & -0.25 & -0.21 \\
        DIN (Base)             & 0.00  & 0.00  \\
        DCNv2                        & +0.11 & +0.19 \\
        APG                          & +0.21 & +0.27 \\
        \midrule
        \multicolumn{3}{l}{\textit{Group 2: Scalable Architectures}} \\
        \midrule
        HiFormer                     & +0.33 & +0.59 \\
        HSTU                         & +0.58 & +0.61 \\
        MTGR                         & +0.64 & +0.69 \\
        HHFT                         & +0.36 & +0.67 \\
        OneTrans                     & \underline{+0.82} & +0.75 \\
        RankMixer (Prod. Base)                  & +0.39 & \underline{+0.77} \\
        \midrule
        \textbf{TmallGS (Ours)}      & \textbf{+1.12} & \textbf{+1.26} \\
        \bottomrule
    \end{tabular}
\end{table}

\textbf{Analysis:} TmallGS consistently outperforms all baselines, revealing two key insights:
(1) \textbf{Compute-Centric Shift:} Scalable architectures (Group 2) generally surpass memory-bound DLRMs (Group 1), confirming that scaling backbone capacity and sequence length ($L=1500$) is essential for capturing complex dependencies.
(2) \textbf{Decoupling Superiority:} While OneTrans improves AUC, its GAUC lags behind RankMixer, indicating \textit{signal dilution} in generic Transformers. TmallGS achieves the highest gains (+1.12\% AUC, +1.26\% GAUC) by utilizing Decoupled FiLM Fusion to preserve high-frequency matching signals, proving that domain-specific decoupling is critical for adapting LLM architectures to search.

\subsection{RQ2: Ablation Studies}
To systematically verify the contribution of each proposed component, we conduct ablation studies on the Tmall Search offline dataset. We create variants by removing ("w/o") or replacing ("$\rightarrow$") specific modules. The results are summarized in Table \ref{tab:ablation}.
\begin{table}[h!]
    \centering
    \caption{Ablation studies of TmallGS. We report the relative performance drop ($\Delta$) in percentage.}
    \label{tab:ablation}
    \renewcommand{\arraystretch}{1} 
    \setlength{\tabcolsep}{6pt}     
    \begin{tabular}{l|cc}
        \toprule
        \textbf{Model Variants} & \textbf{$\Delta$ AUC (\%)} & \textbf{$\Delta$ GAUC (\%)} \\
        \midrule
        \textbf{TmallGS (Full Model)} & \textbf{-} & \textbf{-} \\
        \midrule
        \multicolumn{3}{l}{\textit{A. Interaction Decoupling}} \\

        \hspace{1em}\textit{Setting: w/ Cross-Fea in Block} & & \\

        \hspace{2em} w/o FiLM Fusion           & -0.21 & -0.32 \\
        \hspace{2em} FiLM $\rightarrow$ Concat & -0.15 & -0.24 \\
        \addlinespace[3pt] 
        \hspace{1em}\textit{Setting: w/o Cross-Fea in Block} & & \\
        \hspace{2em} w/o FiLM Fusion           & -0.17 & -0.19 \\
        \hspace{2em} FiLM $\rightarrow$ Concat & -0.10 & -0.11 \\
        \midrule
        \multicolumn{3}{l}{\textit{B. Noise Filtering}} \\
        \hspace{1em} Gating $\rightarrow$ Std. Attn & -0.13 & -0.16 \\
        \midrule
        \multicolumn{3}{l}{\textit{C. Tokenization}} \\
        \hspace{1em} Only DCP   & -0.08 & -0.06 \\
        \hspace{1em} Only FSR   & -0.13 & -0.09 \\
        \hspace{1em} FSR+DCP $\rightarrow$ MLP          & -0.17 & -0.11 \\
        \hspace{1em} FSR+DCP $\rightarrow$ SwiGLUFFN       & -0.05 & -0.03 \\
        \midrule
        \multicolumn{3}{l}{\textit{D. Optimizations \& Training}} \\
        \hspace{1em} w/o Bias Net                   & -0.07 & -0.13 \\
        \hspace{1em} w/o Prog. Train                & -0.11 & -0.08 \\
        \hspace{1em} w/o Pairwise Loss              & +0.05 & -0.87 \\
        \hspace{1em} w/o UQH Seq              & -0.65 & -0.29 \\
        \bottomrule
    \end{tabular}
    \vspace{-0.4cm} 
\end{table}

\textbf{Impact of Structural Designs.}
Block A confirms the necessity of Interaction Decoupling. Removing FiLM Fusion causes the sharpest GAUC drop (-0.32\%), validating that deep Transformers tend to dilute explicit matching signals. FiLM's multiplicative modulation proves superior to additive concatenation (-0.24\%) in recovering these signals.
In Block B, Noise-Adaptive Gating outperforms standard attention (-0.16\% GAUC), effectively mitigating the attention sink problem in noisy behavior sequences.
Block C demonstrates the efficiency of our Tokenization. The FSR+DCP pipeline matches the heavy SwiGLU-FFN in performance (gap <0.03\%) while reducing projection FLOPs by $\sim$33\%, achieving an optimal efficiency-accuracy trade-off.

\textbf{Impact of Optimization Strategies.}
Training protocols are critical for search ranking. Removing the Pairwise Loss causes a catastrophic drop in GAUC (-0.87\%) despite stable AUC, confirming its indispensable role in optimizing intra-session ranking order. Conversely, removing User Query History (UQH) severely impacts global AUC (-0.65\%), underscoring its value for long-term intent modeling. Finally, both the Context-Aware Bias Net and Progressive Training contribute to stability, with their removal leading to consistent degradation.
\begin{table}[t]
    \centering
    \caption{Online A/B Testing Results (30 Days). All metric gains are statistically significant ($p < 0.05$).}
    \label{tab:online_results}
    \setlength{\tabcolsep}{15pt} 
    \renewcommand{\arraystretch}{1.2} 
    \begin{tabular}{l|c}
        \toprule
        \textbf{Metric} & \textbf{TmallGS ($\Delta$)} \\
        \midrule
        PV-AUC   & \textbf{+0.79\%} \\
        Imp-GAUC & \textbf{+0.34\%} \\
        UCTCVR   & \textbf{+1.38\%} \\
        GMV      & \textbf{+1.52\%} \\
        Latency  & \textbf{+6 ms} \\
        \bottomrule
    \end{tabular}
\end{table}
\subsection{RQ3: Scalability Analysis}
\begin{figure}[t]
    \centering

    \begin{minipage}[b]{0.49\linewidth} 
        \centering
        \includegraphics[width=\linewidth]{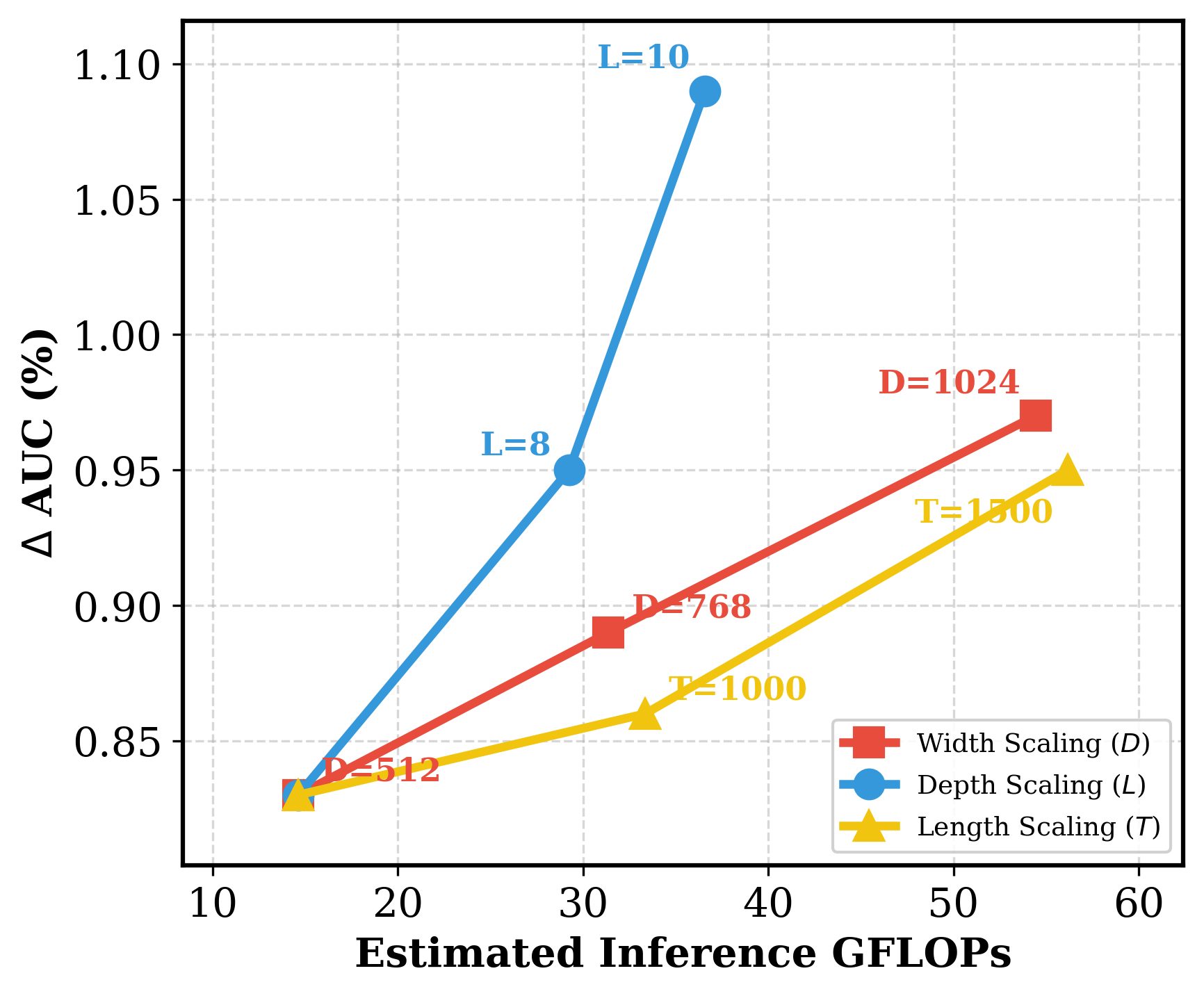}

        \label{fig:scaling_auc}
    \end{minipage}
    \hfill 

    \begin{minipage}[b]{0.49\linewidth}
        \centering
        \includegraphics[width=\linewidth]{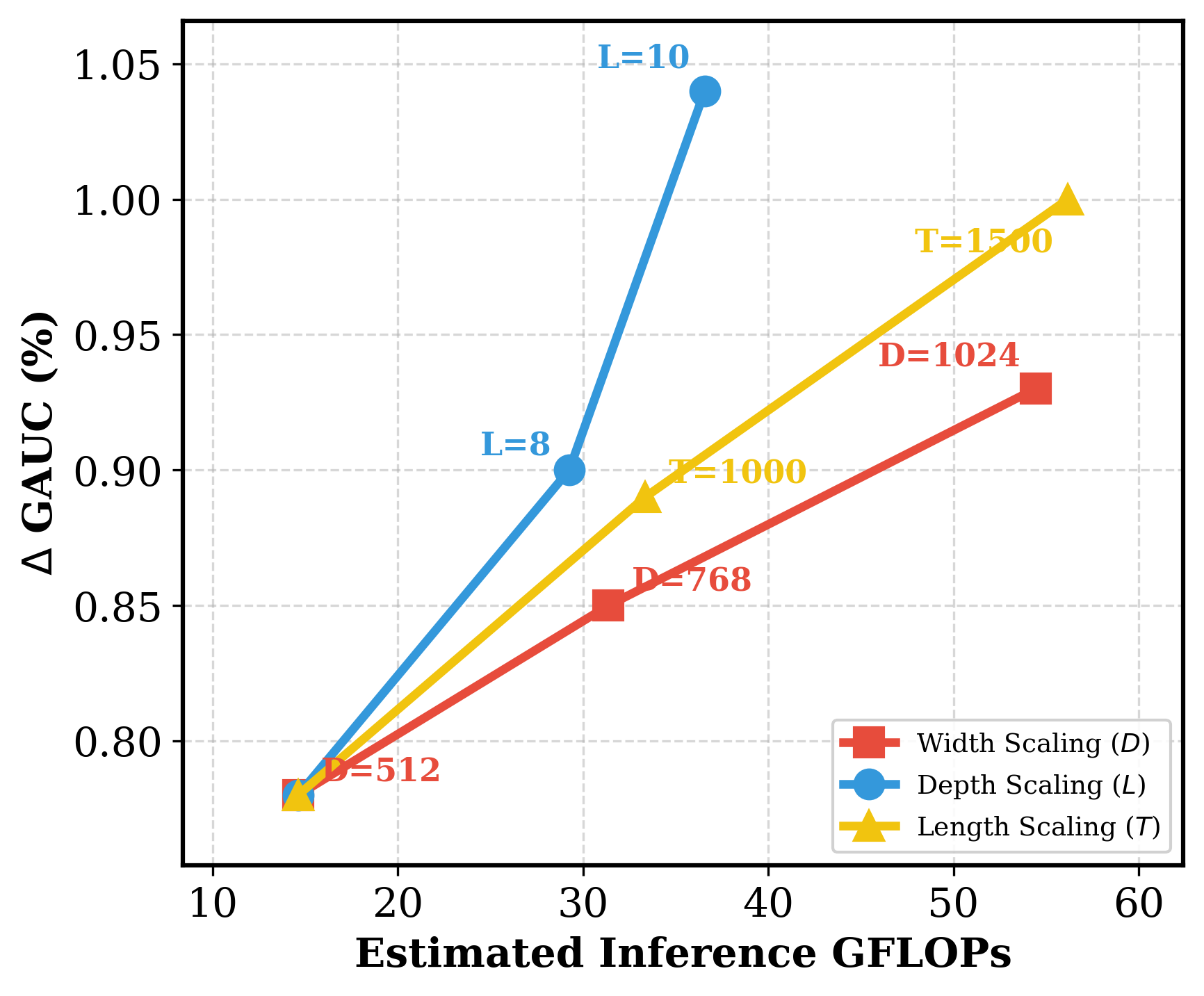}

        \label{fig:scaling_gauc}
    \end{minipage}
    
    \caption{Scaling Laws of TmallGS over DIN baseline. Performance scales monotonically with compute (GFLOPs) across Width, Depth, and Length.}
    \label{fig:scaling_laws}
\end{figure}
We probe the scaling properties of TmallGS against a cold-start DIN baseline along three dimensions: \textbf{Width} ($D \in \{512, 768, 1024\}$), \textbf{Depth} ($L \in \{4, 8, 10\}$), and \textbf{Length} ($T \in \{500, 1000, 1500\}$). 

As shown in Figure \ref{fig:scaling_laws}, TmallGS adheres to scaling laws, exhibiting monotonic performance gains across all dimensions as compute budget increases. Notably, depth scaling yields the steepest trajectory, offering the highest ROI for ranking precision. Furthermore, length scaling maintains robust linear growth up to $T=1500$ without saturation, confirming that our \textit{Noise-Adaptive Gating} effectively prevents noise accumulation in ultra-long sequences.

\subsection{RQ4: Deployment Strategy \& Efficiency}
Migrating from a legacy DLRM to a deep Transformer backbone in a mature industrial system presents unique challenges. In this section, we discuss our solutions for effective warm-up and efficient online serving.
\begin{table}[h!]
    \centering
    \caption{Comparison of Warm-up Strategies.}
    \label{tab:warmup}
    \vspace{-0.4cm} 
    \setlength{\tabcolsep}{10pt}
    \renewcommand{\arraystretch}{1.2}
    \begin{tabular}{l|cc}
        \toprule
        \textbf{Warm-up Strategy} & \textbf{$\Delta$ AUC (\%)} & \textbf{$\Delta$ GAUC (\%)} \\
        \midrule
        Cold Start & - & - \\
        Direct Loading & +0.46 & +0.34 \\
        \textbf{Two-Stage (Ours)} & \textbf{+1.42} & \textbf{+1.09} \\
        \bottomrule
    \end{tabular}
    \vspace{-0.4cm} 
\end{table}
\subsubsection{\textbf{Two-Stage Representation Alignment}}
To effectively inherit massive legacy embeddings, we address the \textit{manifold misalignment} issue where direct loading degrades performance. We propose a \textbf{Two-Stage Warm-up}: first freezing sparse embeddings to align the dense backbone, followed by end-to-end joint fine-tuning. As shown in Table \ref{tab:warmup}, this strategy significantly outperforms direct loading, yielding gains of \textbf{+1.42\% AUC} and \textbf{+1.09\% GAUC}.

\subsubsection{\textbf{Online Configuration}}
Considering the strict latency constraints and the high QPS of Tmall Search, directly deploying a billion-scale LLM is infeasible. Guided by the Scaling Law analysis in RQ3, we prioritize sequence length over model depth, configuring TmallGS with Layers=4, Dimension=1024, Head=8, and a sequence length of 1500. The dense backbone contains approximately \textbf{0.05 billion} parameters. This configuration strikes an optimal balance: the extended sequence length captures long-term user interests, while the shallow-but-wide structure ensures massive parallelism on GPUs, keeping the inference latency within the SLA.

As shown in Table \ref{tab:online_results}, TmallGS achieves remarkable improvements across all core business metrics. The \textbf{+1.52\% lift in GMV} and \textbf{+1.38\% lift in UCTCVR} indicate higher user satisfaction and conversion efficiency. While the average latency increases slightly (+6ms) due to the dense backbone, it remains strictly within the production SLA, proving that TmallGS successfully strikes a favorable balance between high-capacity modeling and system efficiency.

\section{Conclusion}
In this work, we presented \textbf{TmallGS}, a scalable generative ranking architecture that successfully bridges the gap between LLM-style modeling and high-precision industrial search. By introducing a semantic-interaction decoupling paradigm—specifically through Hierarchical Distribution-Calibrated Tokenization and Decoupled FiLM Late Fusion—we effectively resolved the challenges of feature heterogeneity and signal dilution that plague standard Transformer adaptations.
Extensive experiments confirm that TmallGS not only breaks the performance ceiling of traditional DLRMs but also adheres to neural scaling laws, allowing for predictable gains via increased compute. Crucially, successful online deployment in Tmall Search, yielding a \textbf{+1.52\% GMV lift} and \textbf{+1.38\% UCTCVR lift}, validates the commercial viability of compute-intensive backbones under strict latency constraints.
This work marks a definitive paradigm shift from memory-bound to compute-centric ranking in Tmall. For future work, we plan to explore scaling TmallGS to billion-parameter levels and integrating multi-modal generative capabilities to further evolve the next-generation search experience.

\section*{Acknowledgments}
We acknowledge the Alibaba PAI Team for releasing TorchEasyRec~\cite{torcheasyrec2024}, which supported the engineering implementation and large-scale training of our experiments.

\clearpage
\bibliographystyle{ACM-Reference-Format}
\bibliography{sample-base}

\clearpage
\appendix

\clearpage

\section{Detailed Training Algorithm}
Due to space constraints in the main text, we provide the comprehensive training procedure of TmallGS in Algorithm \ref{alg:tmallgs_detailed}. This algorithm details the end-to-end flow, including the Hierarchical Distribution-Calibrated Tokenization, the Field-Adaptive Backbone forward pass, and the Error-Aware Progressive optimization strategy.

\begin{algorithm}
\small 
\caption{Training Procedure of TmallGS (Detailed)}
\label{alg:tmallgs_detailed}

\SetKwInOut{Input}{Input}
\SetKwInOut{Output}{Output}
\SetKwFunction{StopGradient}{StopGradient}

\SetAlCapSkip{1em} 

\Input{Training dataset $\mathcal{S}$, Max layers $L$, Hyperparameters $\lambda, \gamma$}
\Output{Optimized model parameters $\Theta$}

\While{not converged}{
    Sample batch $\mathcal{B} = \{(u, q, \mathcal{C}, \mathbf{y})\} \sim \mathcal{S}$\;
    
    \tcc{1. Hierarchical Dist-Calibrated Tokenization (Sec 4.2)}
    \For{each feature field $i \in \{1, \dots, F\}$}{
        Global Aggregation: $z_i \leftarrow \frac{1}{L \cdot d_i} \sum \mathbf{X}_{i}$\;
    }
    Generate Saliency Weights: $\mathbf{w} \leftarrow \sigma(\mathbf{W}_{ex} \cdot \delta(\mathbf{W}_{sq} \mathbf{z}))$\;
    \For{each feature field $i$}{
        Field-Wise Reweighting: $\tilde{\mathbf{X}}_i \leftarrow w_i \cdot \mathbf{X}_i$\;
        Dist-Calibrated Projection: $\mathbf{E}_i \leftarrow \text{MLP}_{cal}(\tilde{\mathbf{X}}_i)$\;
    }
    
    \tcc{2. Unified Sequence Construction (Sec 4.3)}
    Construct sequence with Context Anchor and Field Tokens:\;
    $\mathbf{E}_{in} \leftarrow [\mathbf{e}_{bias}; \mathbf{e}_{ctx}; \mathbf{e}_{qry}; \mathbf{H}_{uih}; \mathbf{H}_{uqh}; \mathbf{e}_{cand}^{light};]$\;
    
    \tcc{3. Backbone \& Error-Aware Progressive Training}
    Initialize progressive loss $\mathcal{L}_{prog} \leftarrow 0$, loss weight $\alpha^1 \leftarrow 1$\;
    $\mathbf{H}^0 \leftarrow \mathbf{E}_{in}$\;
    
    \For{layer $l = 1$ to $L$}{
        \textit{// Per-Field QKV \& Noise-Adaptive Gating}\;
        $\mathbf{H}^l \leftarrow \text{TmallGS\_Block}(\mathbf{H}^{l-1}, \mathbf{M}_{mask})$\;
        
        Get candidate representation $\mathbf{h}_{c}^l$ from $\mathbf{H}^l$\;
        Auxiliary Prediction: $\hat{y}^l \leftarrow \sigma(\text{AuxHead}^l(\mathbf{h}_{c}^l))$\;
        $\mathcal{L}_{prog} \leftarrow \mathcal{L}_{prog} + \alpha^l \cdot \mathcal{L}_{BCE}(\hat{y}^l, y)$\;
        
        Error Metric: $\delta^l \leftarrow |y - \hat{y}^l|$\;
        $\alpha^{l+1} \leftarrow 1 + \lambda \cdot \StopGradient(\delta^l)$\;
    }
    
    \tcc{4. Decoupled FiLM Fusion \& Bias Net (Sec 4.5 \& 4.6)}
    Extract final states: Candidate $\mathbf{h}_c^L$, Anchor $\mathbf{h}_{bias}^L$\;
    \textbf{Bias Path:} $\text{logit}_{bias} \leftarrow \text{BiasNet}(\mathbf{h}_{bias}^L)$\;
    \textbf{Main Path (FiLM):} 
    Compute $\gamma_c, \beta_c \leftarrow \text{FiLM}(\mathbf{e}_{c}^{all})$\;
    $\mathbf{v}_{final} \leftarrow (1 + \gamma_c) \odot [\mathbf{h}_c^L \oplus \mathbf{e}_c^{all}] + \beta_c$\;
    $\text{logit}_{main} \leftarrow \text{TaskTower}(\mathbf{v}_{final})$\;
    
    Final Prediction: $\hat{y}^L \leftarrow \sigma(\text{logit}_{main} + \text{logit}_{bias})$\;
    
    \tcc{5. Orthogonal Optimization}
    Compute Pairwise Ranking Loss $\mathcal{L}_{pair}$ on $\hat{y}^L$\;
    Total Objective: $\mathcal{L}_{total} \leftarrow \mathcal{L}_{prog} + \gamma \cdot \mathcal{L}_{pair}$\;
    Update $\Theta$ using Split-Optimizer (Adam/AdamW)\;
}
\end{algorithm}

\end{document}